\def\draftversion{false}

\RequirePackage{ifthen}
\ifthenelse{\equal{\draftversion}{true}}{
  \documentclass[aps,prb,11pt,galley,amsmath,amssymb,showpacs,letterpaper]{revtex4}
}{
  \documentclass[aps,prb,10pt,twocolumn,amsmath,amssymb,showpacs,letterpaper,showkeys]{revtex4-1}
}

\usepackage{graphicx}
\usepackage{dcolumn}
\usepackage{amssymb}
\usepackage{bm}
\usepackage{enumitem}

\usepackage{epsfig}
\usepackage{hyperref}

\hypersetup{backref=true, pdfnewwindow=true, colorlinks=true,linkcolor=blue, anchorcolor=blue, citecolor=blue, filecolor=blue, menucolor=blue, urlcolor=blue}

\usepackage[usenames,dvipsnames]{color}

\ifthenelse{\equal{\draftversion}{true}}{
  \marginparwidth 2.7in
  \marginparsep 0.5inhttps://www.overleaf.com/project/62702f1b4f1b03d20db1e6fc
  \newcounter{comm} 
  \def\commnext{\stepcounter{comm}}
  \def\commtext{{\bf\color{blue}[\arabic{comm}]}}
  \def\commmar{{\bf\color{blue}[\arabic{comm}]}}
  \def\dvm#1{\commnext\marginpar{\small DV\commmar: #1}\commtext}
  \def\mtm#1{\commnext\marginpar{\small MT\commmar: #1}\commtext}
  \def\jhm#1{\commnext\marginpar{\small JH\commmar: #1}\commtext}
  \def\vsm#1{\commnext\marginpar{\small VS\commmar: #1}\commtext}
  \def\mlab#1{\marginpar{\small\bf #1}}
  
}{
  \def\dvm#1{}
  \def\mtm#1{}
  \def\jhm#1{}
  \def\vsm#1{}
  \def\mlab#1{}
  
}

\begin{document}

\title{Zig-zag charged domain walls in ferroelectric PbTiO$_3$}

\author{Pavel Marton$^{1,2}$}
\email{marton@fzu.cz}
\author{Mauro A.P. Gonçalves$^{1}$}
\author{Marek Pa\'{s}ciak$^{1}$}
\author{Sabine Körbel$^{1,3}$}
\author{V\v{e}nceslav Chumchal$^{4}$}
\author{Martin Ple\v{s}inger$^{4}$}
\author{Anton\'in Kl\'i\v c$^{1}$}
\author{Jirka Hlinka$^{1}$}
\affiliation{$^1$ Institute of Physics, Czech Academy of Sciences, Na Slovance 2, 182 21 Prague 8, Czech Republic}
\affiliation{$^2$ Institute of Mechatronics and Computer Engineering, Technical University of Liberec, Studentsk\'{a} 2, 461 17 Liberec, Czech Republic}
\affiliation{$^3$ Institute of Condensed Matter Theory and Solid State Optics, Friedrich Schiller University Jena, Max‐Wien‐Platz 1, 07743 Jena, Germany}
\affiliation{$^4$ Department of Mathematics and Didactics of Mathematics, Technical University of Liberec, Studentská 2, 461 17 Liberec, Czech Republic}

\date{\today}

\begin{abstract}

We report a theoretical investigation of a charged 180$^\circ$ domain wall in ferroelectric PbTiO$_3$, compensated by randomly distributed immobile charge defects.
For this we utilize atomistic shell-model simulations and continuous phase-field simulations in the framework of the  Ginzburg-Landau-Devonshire model. 
%
%
We predict that domain walls form a zig-zag pattern and we discuss its properties in a broad interval of compensation-region widths, ranging from a couple to over a hundred nanometers.

\end{abstract}


\keywords{ferroelectric material, PbTiO$_3$, charged domain walls, zig-zag walls, shell-model simulations, phase-field simulations, Ginzburg-Landau-Devonshire model}

\maketitle

\section{Introduction} 

Ferroelectric materials are known to host complicated domain structures that can be used to tune the material properties for a specific application. It turns out that interfaces between the domains -- the domain walls -- have very different properties, e.g., electrical conductivity \cite{guyonnet:2011:conduction,nataf:2020:domain} or phonon modes,\cite{hlinka:2017:terahertz}
from the bulk ferroelectric itself, and therefore understanding the microstructure of the domain wall is receiving an increasing attention in the research of ferroelectrics. 

The domain walls are typically electrically neutral or near-neutral, i.e. $({\bf P}_{1}-{\bf P}_{2})\cdot\textbf{n}\approx0$, where ${\bf P}_{1}$ and ${\bf P}_{2}$ are the ferroelectric polarizations inside the ferroelectric domains on either side of the wall of interest, and ${\bf n}$ is the wall normal. This is because deviation from the charge-neutrality condition leads to a large energy penalty due to the depolarizing electric fields from the polarization-originated bound charges $q_\mathbf{P} = -\mathrm{div}\mathbf{P}$ at the walls. In a perfect dielectric material these depolarization fields would suppress charged interfaces in the early stages of their formation.

However, in spite of apparently unfavorable electrostatics, charged ferroelectric domain walls do exist.\cite{art_denneulin_2022,art_schroder_2012,art_bednyakov_2015,art_jia_2015} This can be rationalized by presence of charged defects such as e.g. electrons, or ionic-type point defects, which compensate for the charge of such walls . The charged walls, as they inevitably involve defects of some kind, are different compared to their neutral counterparts, and therefore exhibit different properties (such as e.g. electric conductivity,\cite{sluka:2013:free} potentially enhanced dielectric response functions \cite{art_li_wu_2012}); this makes them interesting from both fundamental and 
application perspectives.


%

Charged domain walls have been studied theoretically using phase field simulations,\cite{art_ondrejkovic_2013, art_gureev_2012, art_sluka_2012, art_li_wu_2012} density-functional theory,\cite{art_wu_vanderbilt_2006, art_rahmanizadeh_2014, art_sifuna_2020} and analytical considerations and model Hamiltonians. \cite{watanabe:2012:intrinsic,art_zhang_2020} 
In Refs.~\onlinecite{art_wu_vanderbilt_2006, art_rahmanizadeh_2014, art_sifuna_2020} flat walls with heterovalent ions or electrons/holes as compensating charges were modeled. In Refs.~\onlinecite{watanabe:2012:intrinsic,art_zhang_2020} a potential energy functional of charged domain walls was established, whose minimization
leads to zig-zag shaped walls. 
A common feature of these studies is that the compensation charges, if considered at all, are localized in the domain wall, or in a very narrow region close to the wall.
%

It can be expected that in a real situation, the distribution of the compensating charges, especially those that 
do not migrate easily,
may be more complex. In this paper we address a situation with static charges distributed over a wider area. 
We combine shell-model and phase-field simulations to study the microstructure and properties of a domain wall with a compensation region thickness up to a hundred nanometers. This multiscale approach allows us to consider a large system in the phase field simulation while maintaining the accuracy of the atomistic shell model, which is limited to smaller system sizes. We concentrate on the 180$^{\circ}$ tail-to-tail domain wall in PbTiO$_3$ (PTO), a well-known material, for which parameters are available in the literature for both shell-model and phase-field simulations.

In section II we provide details of the utilized methodology, in section III we review the results obtained using shell-model and phase-field simulations, which are in detail discussed in section IV, where we also provide a simplified model explaining the observations. Finally, the paper is concluded in section V. 

\section{Method} 

\begin{figure}
\includegraphics[width=\columnwidth]{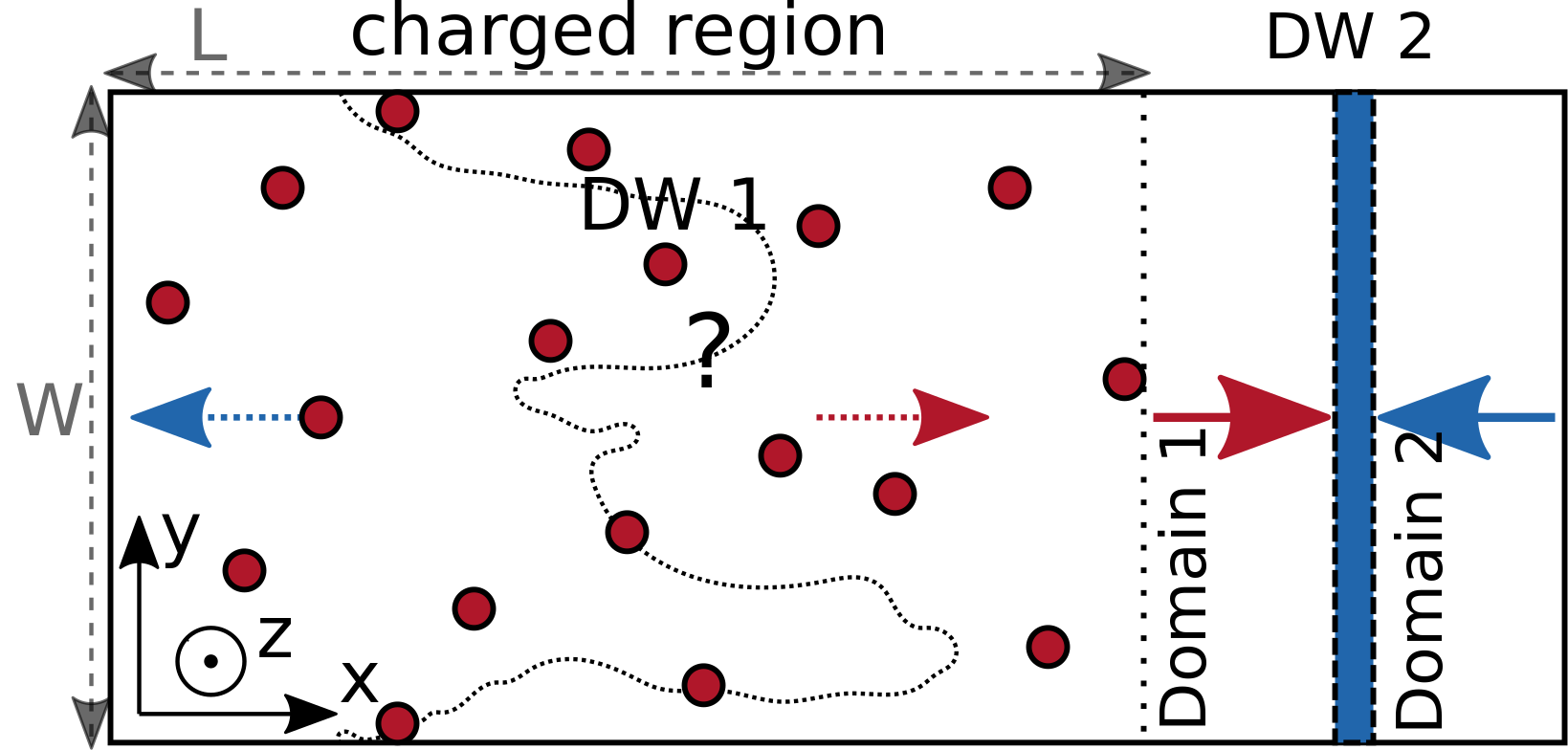}
\caption{Sketch of distribution of point charges in the simulation box. The exact location and properties of the DW1 are not known {\em a priori}. Red dots and blue area correspond to the positive and negative charges, respectively.} 
\label{fig_scheme_dw_1}
\end{figure}

\subsection{Supercell} 

The all-periodic supercell utilized for the calculations is schematically depicted in Fig.\,\ref{fig_scheme_dw_1}. It consists of two domains with spontaneous polarization with opposite directions, separated by two domain walls. 

The negatively charged tail-to-tail wall (DW1) is expected to form within a thick charged layer. As seen in the scheme, the charged region covers most of the supercell volume. The positive charges in this wall have average {\it yz}-planar density  $\sigma_{DW1}=2P_\mathrm{s}$, but they are distributed randomly. The position and shape of the wall DW1 are not restricted in any other way, so it is not a-priori clear whether the wall is going to be planar or will adopt some more complicated profile. Nevertheless, for electrostatic reasons, the wall will be confined to the charged region. 

The positively charged head-to-head wall (DW2) is on the right-hand side of the box and consists of a negatively charged layer, which is chosen to match the charge density caused by the polarization step in the wall, i.e. $\sigma_{DW2}=-2P_\mathrm{s}$.
This wall is not of particular interest in this work and is only present in order to satisfy the periodic boundary conditions. Therefore, it is prepared as thin as possible. Notice that both wall regions are electrically neutral in the sense that the defect charges exactly compensate the polarization-originated bound charge in the walls.

The two charged regions are separated by defect-free areas -- ferroelectric domains, in which the polarization acquires values close to the spontaneous value.

The simulation box is, in general, three-dimensional. While the $x$ dimensions of  Domain 1, Domain 2, and DW2 remain constant in all simulations, we alter the thickness $L$ of the charged region and the height of the simulation box $W$.
%

\subsection{Shell-model setup} 


For atomistic simulations we used a shell model 
potential\cite{Tinte_2004} for $(1-x)$PbMg$_{1/3}$Nb$_{2/3}$O$_3$–$x$PbTiO$_3$
with parameters fitted to first-principles results~\cite{Sepliarsky2011}.
Therein the atomic charges (core + shell) are treated as parameters and attain the following values: Pb$^{+1.80}$, Ti$^{+2.88}$, O$^{-1.56}$, Mg$^{+2.36}$, Nb$^{+3.15}$.
The spontaneous polarization obtained for pure PTO using this model is $P_\mathrm{s}= 0.66$\,Cm$^{-2}$.

The model allows us to study the simulation box described above (Fig.\,\ref{fig_scheme_dw_1}), with the DW1 layer ($L = 48$ unit cells) having a fraction of Ti atoms randomly substituted by Nb and DW2 layer (2 unit cells thick) with all Ti atoms replaced by Mg.
The two unperturbed domains are 7 unit cells thick and the total size of the supercell is 64$\times$44$\times$6 unit cells.

To optimize atomic positions 
we run 
molecular dynamics simulations at low-temperature (T=1K) using the DLPOLY software.\cite{Todorov2006} 
The time step was 0.4 fs. The atoms in the initial configuration were in their ideal cubic positions. After 
equilibration for 30 ps, 
a trajectory of 10~ps was collected and used for calculation of various properties of interest.



Finally, in the simulations the system is mechanically free in all directions, which implies that all dimensions of the supercell can vary during the simulation time. To understand the implications of this approach, 
we also performed calculations imposing the tetragonal strain of PTO (bigger lattice parameter along x). When comparing the results of the two approaches it is evident that 
constraining the lattice parameters has only a minor impact on the results.

\subsection{Phase-field setup} 

The phase-field simulations are performed using the code Ferrodo2,\cite{art_marton_hlinka_2006} which implements the evolution of the ferroelectric polarization in the framework of the Ginzburg-Landau-Devonshire model together with the Landau-Chalatnikov dynamics. Landau, gradient, elastic, electrostrictive and electrostatic interactions are taken into account. For the exact form of the individual energy terms, see Ref.\,\onlinecite{boo_ondrejkovic_marton_2020}. 

The utilized parametrization\footnote{ Full parametrization of the Ginzburg-Landau-Devonshire model utilized here is:  
$\alpha_{1}=-5.42859\times10^{8}$\,JmC$^{-2}$, 
$\alpha_{11}=4.78993\times10^{8}$\,Jm$^{5}$C$^{-4}$, 
$\alpha_{12}=1.13718\times10^{9}$\,Jm$^{5}$C$^{-4}$, 
$\alpha_{111}=-5.47443\times10^{7}$\,Jm$^{9}$C$^{-6}$,  
$\alpha_{112}=1.44549\times10^{7}$\,Jm$^{9}$C$^{-6}$, 
$\alpha_{123}=-5.87811\times10^{8}$\,Jm$^{9}$C$^{-6}$,  
$\alpha_{1111}=-571565$\,Jm$^{13}$C$^{-8}$,  
$\alpha_{1112}=6.04576\times10^{7}$\,Jm$^{13}$C$^{-8}$, 
$\alpha_{1122}=-1.78503\times10^{8}$\,Jm$^{13}$C$^{-8}$, $\alpha_{1123}=-1.59616\times10^{8}$\,Jm$^{13}$C$^{-8}$,
$G_{11}=1\times10^{-10}$\,Jm$^{3}$C$^{-2}$,
$G_{12}=-1\times10^{-10}$\,Jm$^{3}$C$^{-2}$,
$G_{44}=1\times10^{-10}$\,Jm$^{3}$C$^{-2}$,
$C_{11}=3.26537\times10^{11}$\,Jm$^{-3}$,
$C_{12}=9.96364\times10^{10}$\,Jm$^{-3}$,
$C_{44}=6.96305\times10^{10}$\,Jm$^{-3}$,
$q_{11}=1.53698\times10^{10}$\,JmC$^{-2}$,
$q_{12}=1.59913\times10^{9}$\,JmC$^{-2}$,
$q_{23}=5.79024\times10^{9}$\,JmC$^{-2}$,
$\varepsilon_{B}=1$.} of the local part of energy functional for PTO is based on first-principles calculations and was obtained using the same procedure as described in Ref.\,\onlinecite{art_marton_klic_2017}. Thus, the simulation temperature is 0\,K, similar to the shell-model approach. The gradient interaction is chosen as isotropic. The spontaneous polarization resulting from this parametrization is $P_\mathrm{s}= 0.80$\,Cm$^{-2}$ (notice that the spontaneous value differs from the shell-model). We tested that the outcomes of the presented simulations do not particularly depend on the used parametrization and temperature, e.g. using the temperature-dependent Model~I for PTO from Tab.\,4.6 in Ref.\,\onlinecite{boo_ondrejkovic_marton_2020} (@T=298\,K).

The defect charges are represented by point charges (thus, in contrast to the shell model they have no relation to a specific atomic species). For the sake of analytical derivations (see below) we also consider a homogeneous distribution of the compensation charge, which allows us to eliminate local effects related to the discrete nature of point charges. The volume density of compensation charge is chosen as $2P_\mathrm{s}/L$. When integrated along the $x$-axis, it leads to the expected charge density $\sigma_{DW1}=2P_\mathrm{s}$ in the $yz$ plane. 

The spatial step $\Delta$ is chosen similarly to the lattice constant of PTO, $\Delta=0.4$\,nm. We vary the thickness $L$ of the compensation-charge region at DW1; domains 1 and 2 have always thicknesses of 20$\Delta$, DW2 is always 2$\Delta$ thick. 

\begin{figure}
\begin{tabular}{c}
\includegraphics[width=\columnwidth]{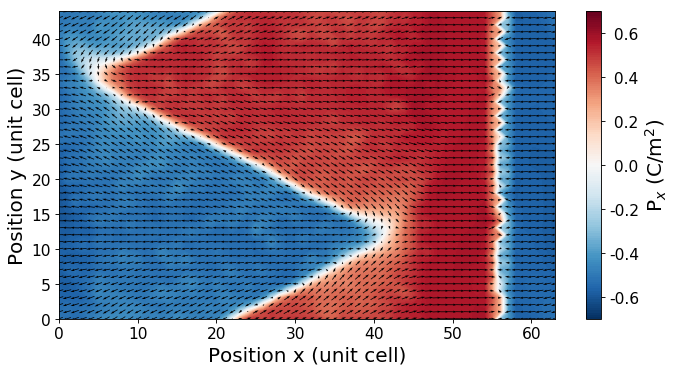}\\
\includegraphics[width=\columnwidth]{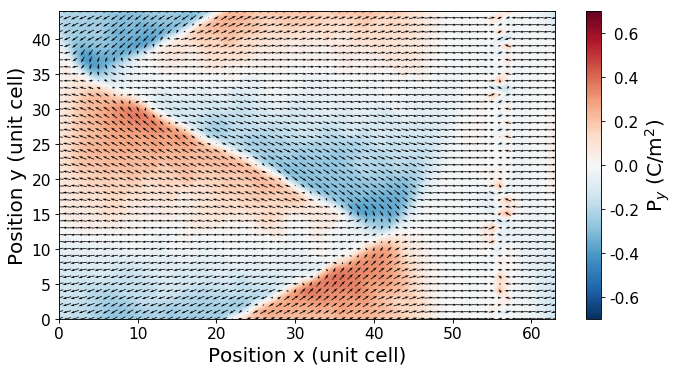}\\
\includegraphics[width=\columnwidth]{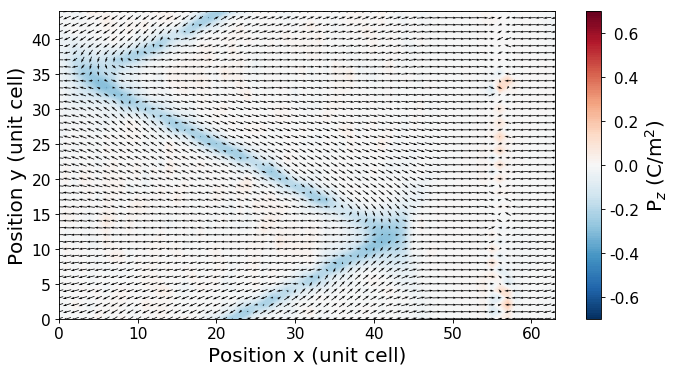}
\end{tabular}
\caption{A zig-zag domain wall as obtained from shell-model simulations. Arrows stand for the polarization of individual unit cells projected to the {\it xy} plane, color represents P$\mathrm{x}$, P$\mathrm{y}$, and P$\mathrm{z}$ components of polarization in individual panels. The domain wall in the figure was obtained for the L=48 and W=45 unit cells.
}
\label{fig_shell_model_pyramid_mauro1}
\end{figure}

The initial ferroelectric polarization was set to almost zero except for a small perturbation, which allows the wall to depart from the strictly planar configuration, corresponding to an unstable, but symmetry-locked state.

The system is mechanically free in all directions.\footnote{The ordinary strategy would be to fix the strain components $e_\mathrm{yy}$, $e_\mathrm{zz}$, and $e_\mathrm{yz}$ to the spontaneous values.} This choice allows us to be consistent with the shell model, where it is difficult to combine mechanical clamping and pressure for different components of the stress-strain boundary conditions. It was tested that the use of such combined boundary conditions has only a minor impact on the results: the polarization relaxes to values relatively close to the spontaneous one in the major part of the simulation box.

The energy functional utilizes the elimination of mechanical strain under the condition of mechanical equilibrium, i.e. the local strains immediately follow the polarization.\cite{art_nambu_sagala_1994} Let us stress that in the two-dimensional simulations all the polarization and strain are treated as fully three-dimensional. By the dimension of simulations we understand here the number of directions in which a sizeable variation of polarization can develop.

\section{Results} 

\subsection{Shell-model simulations} 

Perhaps the most important observation obtained from shell-model simulations is that the domain wall develops into a zig-zag wave inside the charged region,
as depicted in the Fig.\,\ref{fig_shell_model_pyramid_mauro1}(a). The domain wall itself (the transition region where the polarization reverses) is rather narrow irrespective of the thickness of the charged region: the polarization changes its orientation from negative to positive values within a couple of unit cells only.
The zig-zag triangles are not exactly symmetric, they are slightly skewed and has a fading, smoke-like feature near the top. 
The reasons for this deviation from ideal symmetric triangles will be addressed later.

Despite the profound difference between the regions without extra charge (Domains 1,2) and the charged region, we observe that the ferroelectric polarization exhibits no marked change at their interface. Instead, the domain penetrates into the charged region, with the P$_\mathrm{x}$ component close to the spontaneous value. 

Fig.\,\ref{fig_shell_model_pyramid_mauro1}(b) shows the $P_\mathrm{y}$. It is evident that this component of the polarization develops mainly in the triangular domains inside the charged regions. Its magnitude changes linearly along the $y$-direction, going from negative to positive, and abruptly changes sign at the domain wall.

Fig.\,\ref{fig_shell_model_pyramid_mauro1}(c) shows $P_\mathrm{z}$, which is zero everywhere except for the domain wall, where both $P_\mathrm{x}$ and $P_\mathrm{y}$ are close to zero and instead $P_z$ is finite.

\begin{figure*}
\includegraphics[width=2.06\columnwidth]{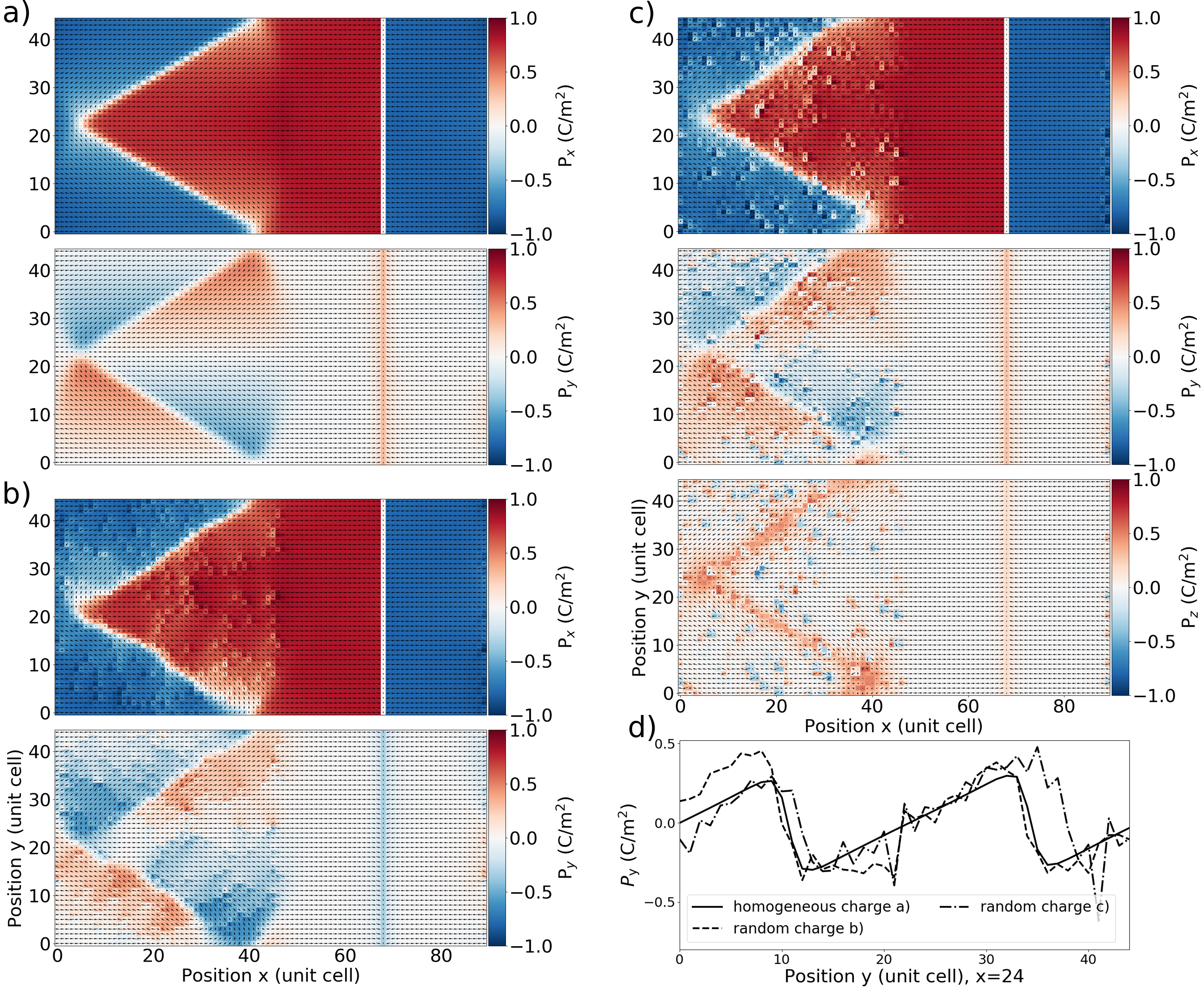}
\caption{A zig-zag domain wall obtained from phase-field simulations. Three types of charge distributions were considered, all of them leading to the same average charge density in the wall.
a) homogeneous charge,
b) charges as in shell-model calculations,
c) charges equal to the elementary charge. Color
represents $P_\mathrm{x}$ and $P_\mathrm{y}$ in the first
and second panel for each case (and $P_\mathrm{z}$ in the third panel of c). d) $P_\mathrm{y}(y)$ evaluated for $x=24$ shows linear dependence inside the triangular domain for all three studied cases, and a steep decrease within the domain wall.}
\label{fig_phase-field_diffent_charges}
\end{figure*}

\subsection{Phase-field simulations} 

The results obtained using the phase-field method are shown in Fig.\,\ref{fig_phase-field_diffent_charges}. 
The dimensions $L$ and $W$ of the charged region are identical to those considered in the shell-model simulations (Fig.\,\ref{fig_shell_model_pyramid_mauro1}). 
The different overall dimensions of the {\it xy}-plane 
plots is caused by different thicknesses of the defect-free domain regions: they are broader in the phase-field simulations than in the shell-model simulations. Three different ways of including the defect charges are considered. Figure\,\ref{fig_phase-field_diffent_charges}(a) was obtained for a homogeneous distribution of the compensating charge and leads to a symmetric zig-zag domains. The compensating charges in Fig.\,\ref{fig_phase-field_diffent_charges}(b) are analogous to the compensating point charges used in the shell model. In Fig.\,\ref{fig_phase-field_diffent_charges}(c) we used fewer point charges with a larger charge (the elementary charge $|e|$).\footnote{The simulation in Fig.\,\ref{fig_phase-field_diffent_charges}(c) was conducted with $z$-dimension of the simulation box equal to 10, i.e. with fully three-dimensional box in order to allow for completely random distribution of charges and avoid formation of chains of charges, as it inevitably happens for two-dimensional simulations.} Notice that the color scale slightly differs between Figs.\,\ref{fig_shell_model_pyramid_mauro1} and \ref{fig_phase-field_diffent_charges}, as the spontaneous polarization predicted by either model is different.


In accordance with the shell-model simulations, the dependence of the $P_\mathrm{y}$ on $y$ within the triangular domain is approximately linear, as can be deduced from the corresponding panels of Fig.\,\ref{fig_phase-field_diffent_charges}(a)-(c). To demonstrate this more clearly, we plot the dependence of  $P_\mathrm{y}$ along the $y$-axis for all three studied charge distributions in Fig.\,\ref{fig_phase-field_diffent_charges}(d). Clearly visible is the linear dependence of $P_\mathrm{y}$ inside the triangular domain, and its rapid decrease in the region of the domain wall.  Notice that $P_\mathrm{y}=0$ on the axis of the triangular domain.
 

The third panel in Fig.\,\ref{fig_phase-field_diffent_charges}(c) displays the $P_\mathrm{z}$ component of the ferroelectric polarization. We observe an out-of-plane component in the region of the DW1, in agreement with the shell-model simulations (see the bottom panel of Fig.\,\ref{fig_shell_model_pyramid_mauro1}).\footnote{The same observation is made for other phase-field simulations presented here, but it is not explicitly visualized.} A footprint of the tendency to form an extra component of polarization in the domain wall can be also observed in  Fig.\,\ref{fig_phase-field_diffent_charges}(a)-(c) for the $P_\mathrm{y}$ component in the DW2 (for the head-to-head 180$^\circ$ DW2 the $y$- and $z$-directions are practically equivalent). The appearance of $P_\mathrm{z}$ in the charged wall is supported by auxiliary density-functional-theory calculations; its magnitude, nevertheless, appears quite sensitive to the details of the defect-charge representation in the narrow walls considered within the relatively small supercells accessible with the first-principles calculations.

Finally, all simulations predict that the polarization vectors in the defect-free regions of Domain~1 and Domain~2 stay close to spontaneous in $x$-direction, justifying the use of rather thin layers to represent them; this allows for a larger portion of the simulation box to be devoted to the region of interest. The wall itself is thicker than a non-charged 180$^\circ$ domain wall.\cite{art_meyer_vanderbilt_2002} In Fig.\,\ref{fig_shell_model_pyramid_mauro1}(b) we observe similar features close to the apex of the triangle as in the shell-model result in Fig.\,\ref{fig_shell_model_pyramid_mauro1}.

In general, there is a very good agreement between the shell-model and phase-field simulations. Both predict a zig-zag wall within the charged region.

\section{Discussion} 

\subsection{One-dimensional character of domain-wall modulation} 

The simulations presented here are mostly two-dimensional (in the {\it xy} plane ). Simulations using three-dimensional simulation boxes were conducted as well: they systematically predict the development of a zig-zag wall profile contained within the charged layer and with the modulation direction along either {\it y} or {\it z} axis. Even though the modulation direction was systematically observed to align with a pseudo-cubic direction, this does not need to be the case for all materials: the anisotropy of the transversal permittivity for the spontaneous state and the concrete form of the short-range interaction may lead to an alternative preferred wall-modulation direction, see the discussion below. Here we conclude that the two-dimensional simulations satisfactorily represent the 3D wall behavior. 

\subsection{Origin of the zig-zag profile of the wall} 

For the sake of gaining insight into the zig-zag character of the wall, it is instrumental to use the simplified homogeneous-charge approximation, already used e.g. in Fig.\,\ref{fig_phase-field_diffent_charges}(a). It allows us to disregard the actual positions of the defects while pertaining their average effect.

There is a simple reason why the domain wall cannot stay flat: in order for the bound-charge density due to the variation of solely $P_\mathrm{x}$ (recall that $\rho_{P} = -\mathrm{div}\mathbf{P}$) to approximately compensate the defect charge, the change in $P_\mathrm{x}$ between its spontaneous values would need to be approximately linear. Such a strong deviation from the spontaneous polarization in such a large volume region would cost too much energy.


\begin{figure}
\includegraphics[width=\columnwidth]{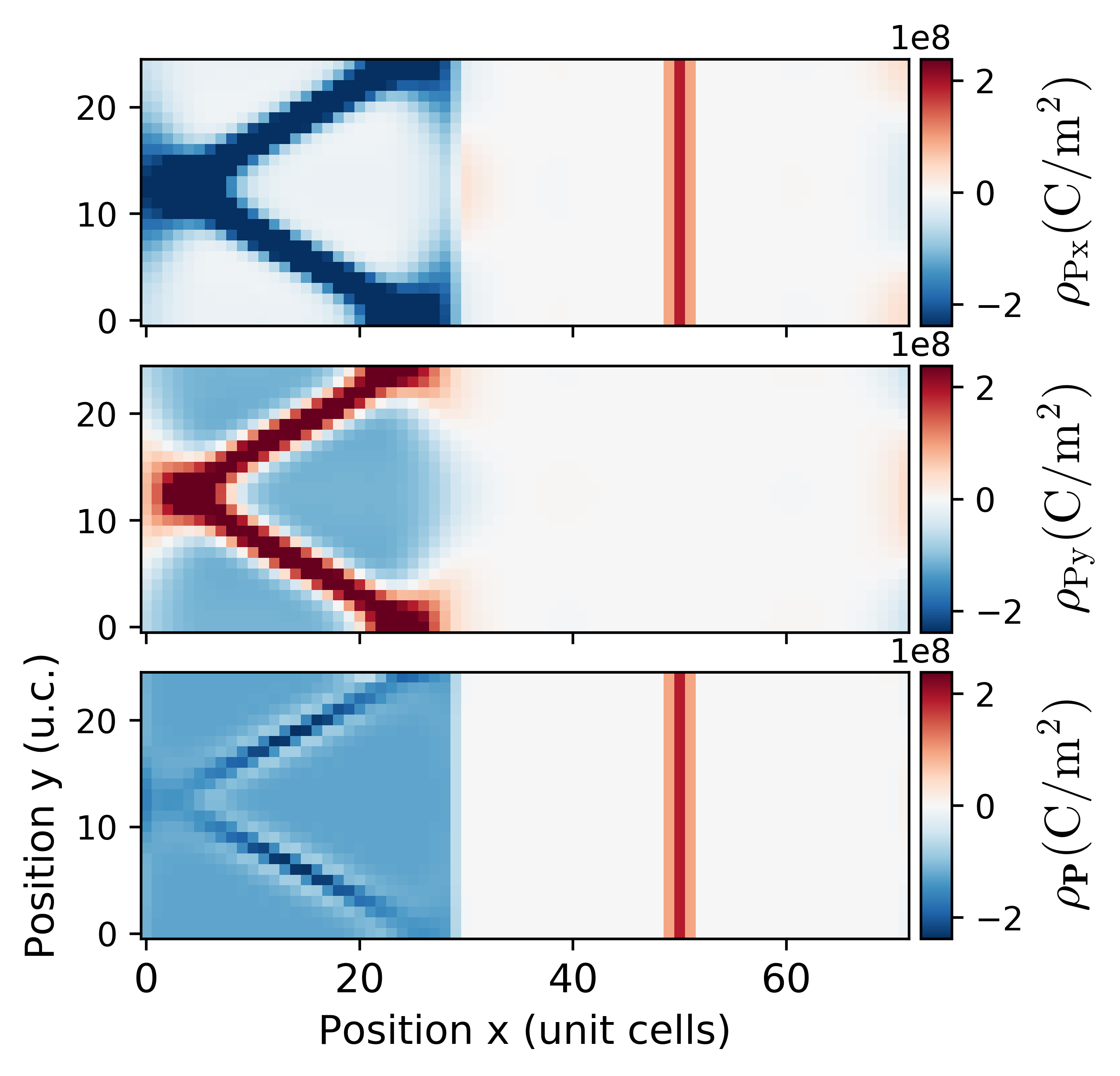}
\caption{Polarization-originated bound charges, stemming from the $P_\mathrm{x}$ (upper), $P_\mathrm{y}$ (middle) components of polarization, and their sum (bottom). For better visibility of the the DW1 region we artificially reduce the intensity of charge in the DW2 by a factor of  10.}
\label{fig_script_plot_divergence}
\end{figure}

In order to avoid the paraelectric state, the $P_\mathrm{y}$ component of polarization develops. A closer look at the polarization in Figs.\,\ref{fig_shell_model_pyramid_mauro1}, \ref{fig_phase-field_diffent_charges} reveals that the $P_\mathrm{y}=0$ at the axis of the triangle. 
The $P_\mathrm{y}$ increases linearly along the direction of the {\it y}-axis, and sharply drops in the region of the zig-zag domain wall, thus being zero on average (see also Fig.\,\ref{fig_phase-field_diffent_charges}(d)).

The visualization of how the ferroelectric polarization gives rise to bound charges $\rho_\mathbf{P}= -\mathrm{div} \mathbf{P} =-\left( \frac{\partial P_\mathrm{x}}{\partial x} + \frac{\partial P_\mathrm{y}}{\partial y} + \frac{\partial P_\mathrm{z}}{\partial z}\right) = \rho_\mathrm{Px}+\rho_\mathrm{Py}+\rho_\mathrm{Pz}$ is depicted in Fig.\,\ref{fig_script_plot_divergence} for the case with relatively small $L$, for which the effect is more pronounced. The upper panel shows the contribution to the polarization-originated charge from $P_\mathrm{x}$. It is negative  in the region of the zig-zag wall (as it needs to be in tail-to-tail wall), and zero elsewhere, i.e. $P_\mathrm{x}$ stays almost constant within the triangles. The middle panel depicts the contribution of the $P_\mathrm{y}$ component. Its linear increase along the $y$-axis leads to a negative charge (blue in the triangular domain), while the sharp drop in the wall results in a large positive charge (red at the triangle edges). Adding these two\footnote{Notice that $\rho_\mathrm{Pz}$ is zero in two-dimensional simulations.} contributions together (bottom panel) results in an almost constant negative charge within the entire charged region, which matches the positive charge due to immobile defects. Notice that the $\rho_\mathbf{P}$ is not exactly constant in the wall region, but even there the differences (of lighter and darker areas) approximately cancel and lead to local electric fields only, which are energetically acceptable. 

Thus, while the $P_\mathrm{x}$ component is responsible for the negative charge of the wall, $P_\mathrm{y}$ produces a positive and negative regions, which approximately compensates the negative wall and all the positive defect charges at the background. Uncompensated charges are largely suppressed and there are almost no sources of energetically-costly electric fields.

Notice that the deviation of $P_\mathrm{y}$ costs energy; however, as this is a transversal deviation from spontaneous vector, it costs less energy than the longitudinal deviation. This effect is related to the usually larger permittivity in the transversal than in the longitudinal direction in ferroelectrics.\cite{art_sluka_2012,art_ondrejkovic_2013}

To return back to the original question about the reason why the shape of the wall is zig-zag. It is the shortest interface, which at the same time penetrates the whole charged layer. In other words, it is the energetically most economic wall from the perspective of the domain-wall surface energy density, which allows for the charge compensation mediated by variation of  $P_\mathrm{y}$.

A similar mechanism can be expected in cases where we deal with point charges instead of a homogeneous charge density: the variation of the electric field due to local charges averages out due to the large number of involved defects. Figure\,\ref{fig_phase-field_diffent_charges}(c) demonstrates the robustness of the zig-zag arrangement, which is present even for strongly charged point defects.

\subsection{Natural shape of triangular domains} 
\label{sec_w_natural}


%
%

\begin{figure}
\begin{tabular}{c}
\includegraphics[width=0.91\columnwidth]{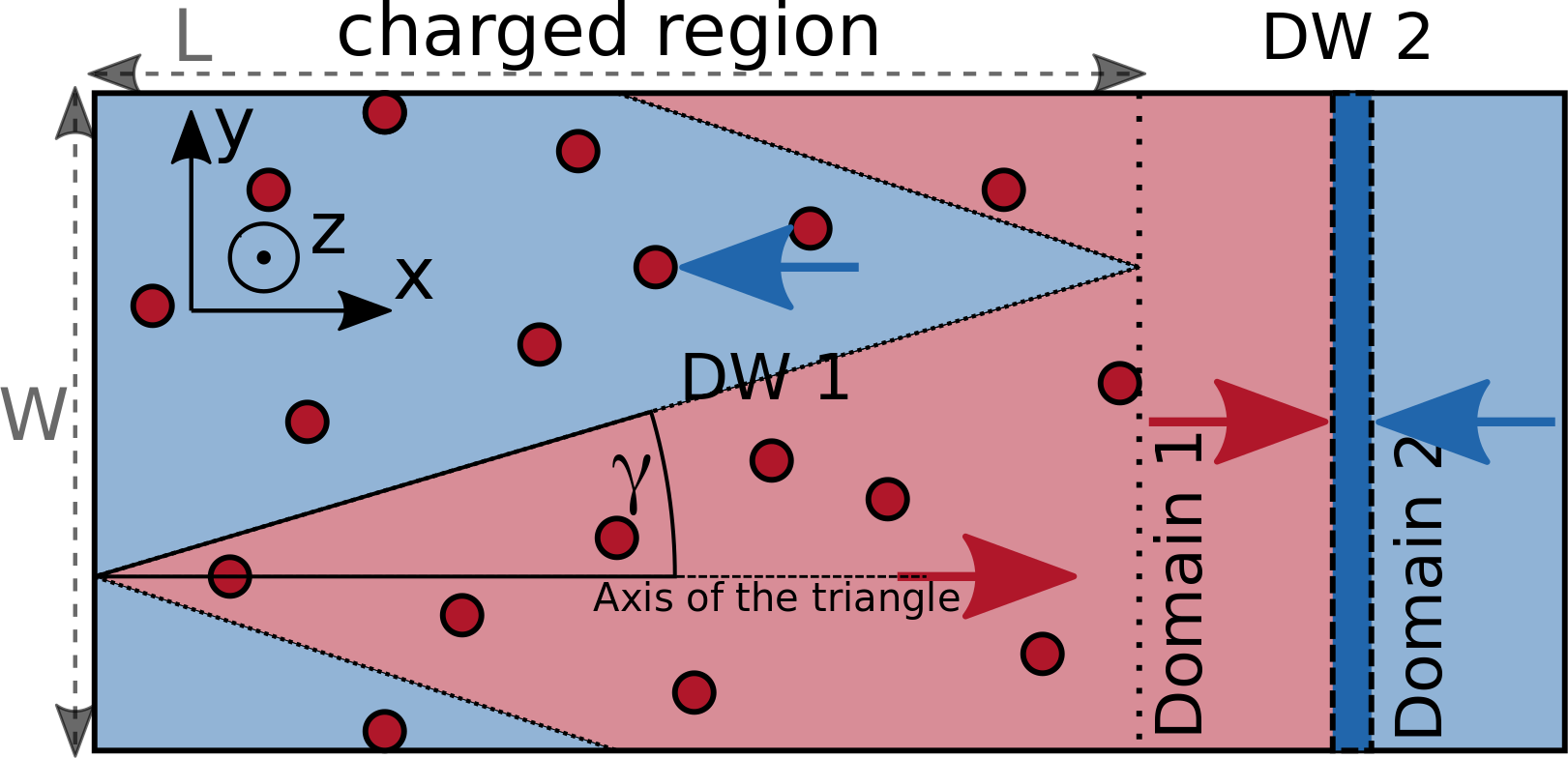}\\
\includegraphics[width=\columnwidth]{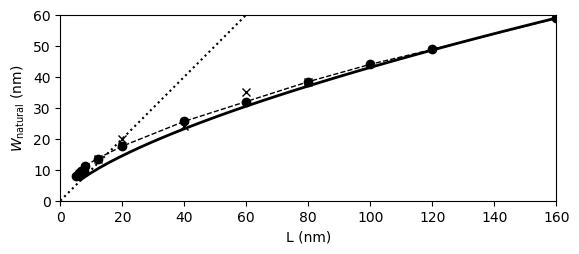}
\end{tabular}
\caption{Top: schematic picture of the zig-zag pattern, which is used in the derivation of the W$_\mathrm{natural}$(L).
Bottom: dependence of W$_\mathrm{natural}$(L). Bullets: phase-field simulations with homogeneous charge, dashed line is just a connection of these. 
Crosses: phase-field simulations with randomly distributed point charges, as in Fig.\,\ref{fig_phase-field_diffent_charges}(b).
Solid line: simplified model.
The dotted line indicates dependence W=L, i.e., a triangles with identical width and height.}
\label{fig_W_natural}
\end{figure}

So far, the dimensions $W$ for the given $L$ was chosen in such a way that it produces a more or less ''acceptable'' modulation length for the zig-zag wall (i.e. a single triangle develops for each domain in the simulations).
In the following we investigate how the energetically optimal base of the triangle $W_\mathrm{natural}$ depends on the width of the compensating charge distribution $L$.

We have used phase-field simulations to visualize the dependence in Fig.\,\ref{fig_W_natural}. For the sake of the plot, the $W_\mathrm{natural}$ for each considered $L$ (bullets  and crosses in the figure) was evaluated as the $W$, for which the optimized configuration with one triangle in the simulation box has the lowest planar energy density in the {\it yz} plane. Thus, $W_\mathrm{natural}$ is the zig-zag period, which would materialize as energetically most favourable without constraints on $W$ imposed by the $y$-dimension of the simulation box.

To understand the behavior of $W_\mathrm{natural}(L)$ in Fig.\,\ref{fig_W_natural}, we use here again the homogeneous-defect-charge approximation, and the strictly triangular zig-zag domain wall, as depicted in
Fig.\,\ref{fig_W_natural}(top panel).
The first contribution to the planar energy density of the wall per unit area in the {\it yz} plane, $\mathcal{E}_1$, is the surface energy of the wall as a function of $W$ and $L$
\begin{eqnarray}
\mathcal{E}_1=2\mu\sqrt{\left(\frac{L}{W}\right)^2+\frac{1}{4}}~,
\label{eqn_E1}
\end{eqnarray}
where $\mu$ represents the surface energy density of the wall. $\mathcal{E}_1$ grows with decreasing of $W$, because narrow triangular domains have a (relatively) large wall surface. 

The second contribution to the energy density, $\mathcal{E}_2$,  accounts for the deviation of the polarization vector in the $y$-direction from zero by $\Delta P_\mathrm{y}(\mathbf{r})$. Let us assume that i) the deviations $\Delta P_\mathrm{y}$ are small enough and the Landau energy density $f_\mathrm{L}$ can be considered to be quadratic 
\begin{eqnarray}
f_\mathrm{L}=f_\mathrm{L}(\mathbf{P}_\mathrm{s})+\alpha_1^\prime (\Delta P_\mathrm{y})^2~,
\end{eqnarray}
ii) $\Delta \mathbf{P}_\mathrm{y}(\mathbf{r})$ linearly increases along $y$ while P$_\mathrm{x}$ remains constant, iii) the domain wall has zero thickness, and iv) all other energy contributions, e.g. due to electric fields originating in incomplete compensation of the defect charge by the polarization variation can be neglected. Under these assumptions, it can be derived that
\begin{eqnarray}
\label{eqn_E2}
\mathcal{E}_2
&=&\frac{2}{W}\int_\mathrm{T}f_\mathrm{L}\left(P_\mathrm{sx},\Delta P_\mathrm{y},0\right) - f_\mathrm{L}(\mathbf{P}_\mathrm{s})  d\mathbf{r}\nonumber\\
&=&\frac{4}{W}\int_{0}^{L}\left(\int_0^{x \mathrm{tg}(\gamma)} \frac{1}{2\chi_{\perp}}\left(\frac{2P_\mathrm{s}y}{L}\right)^2 dy\right)dx\nonumber\\
&=&\frac{P_\mathrm{s}^2 W^2}{12 \chi_\perp L}~.
\end{eqnarray}
Here $T$ is the area of a single triangle with the base $W$, $\chi_{\perp}=1/(2\alpha_1^\prime)$ represents the transversal susceptibility in the spontaneous ferroelectric state $\mathbf{P}_\mathrm{s}=(P_\mathrm{sx},0,0)$, and $2\gamma$ is the angle at the triangle apex. In the derivation we use the linear dependence of $\Delta P_\mathrm{y}$ on $y$ (with zero on the axis of the wedge)
\begin{eqnarray}
\Delta P_\mathrm{y}=\frac{2P_\mathrm{s}}{L}y~,
\end{eqnarray}
which allows for exact compensation of the homogeneous background charge (the coefficient of proportionality is exactly equal to the homogeneous charge density in the charged layer). Therefore, the maximal $\Delta P_\mathrm{y}$ found in the triangle is proportional to the width $W$ of the triangle. From Eqn.\,\ref{eqn_E2} it follows that $\mathcal{E}_2$ grows with increasing $W$, as sharper triangles require smaller maximal values of $\Delta P_\mathrm{y}$. 

The total energy density of the wall per unit area in the {\it yz} plane is $\mathcal{E}=\mathcal{E}_1+\mathcal{E}_2$.  $W_\mathrm{natural}$ is the $W$ for which the energy is minimal, requiring $d\mathcal{E}/dW=0$. Tee complex nature of the $\mathcal{E}$ does not allow to express the $W_\mathrm{natural}(L)$ analytically in a simple form; for large $L$ the solution asymptotically approaches
\begin{eqnarray}
W_\mathrm{natural}(L)=\left(\frac{12 \mu \chi_\perp L^2 }{P_\mathrm{s}^2}\right)^{\frac{1}{3}}~.
\end{eqnarray}
The numerical solution is plotted in Fig.\,\ref{fig_W_natural} (solid line), taking into account $\mu=350$\,mJ/m$^2$, and $\chi_\perp=279\varepsilon_0$ evaluated from the utilized Landau potential $f_\mathrm{L}(\mathbf{P})$.\footnote{The $f_\mathrm{L}(P_\mathrm{sx},P_\mathrm{y},0)$= 1.65687$\times 10^8$ + 2.02795$\times 10^8 P_\mathrm{y}^2$ + 4.15809$\times 10^8 P_\mathrm{y}^4$ - 
 1.62438$\times 10^7 P_\mathrm{y}^6$ - 5.71570$\times 10^5 P_\mathrm{y}^8$.} 

For large $L$ the polarization will be close to the spontaneous one within the entire area of each triangle, and the adopted assumptions are largely valid, and the prediction of the simplified model agrees well with the numerical data. On the other hand, for small $L$ we see a discrepancy between the numerical data and the predicted $W_\mathrm{natural}(L)$. This is because the $\Delta \mathbf{P}_\mathrm{y}(\mathbf{r})$ can no longer be considered small (see e.g. Fig.\,\ref{fig_shell_model_pyramid_mauro1}), the volume of the wall itself becomes significant, and the adopted assumptions, in particular that of a constant $P_\mathrm{x}$, are no longer satisfied. This will lead to smaller $\mathcal{E}_2$, and thus to broader triangles obtained from simulations then it is predicted by the theory, which indeed we observe in Fig.\,\ref{fig_W_natural}.

In general, a tendency to form narrow triangles (small $\gamma$) can be expected in materials with small planar energy density of the domain wall $\mu$ (small cost of the wall area), small spontaneous polarization magnitude $P_\mathrm{s}$ (small defect-charge volume density), small susceptibility $\chi_\perp$ (large energy cost due to induced transversal polarization $\Delta \mathbf{P}_\mathrm{y}(\mathbf{r})$), and small background permittivity $\varepsilon_\mathrm{B}$ (reflecting contributions to the permittivity due to high-frequency polar modes and electronic degrees of freedom). 

\subsection{Deviations from the ideal zig-zag pattern} 

\begin{figure}
\includegraphics[width=1.08\columnwidth]{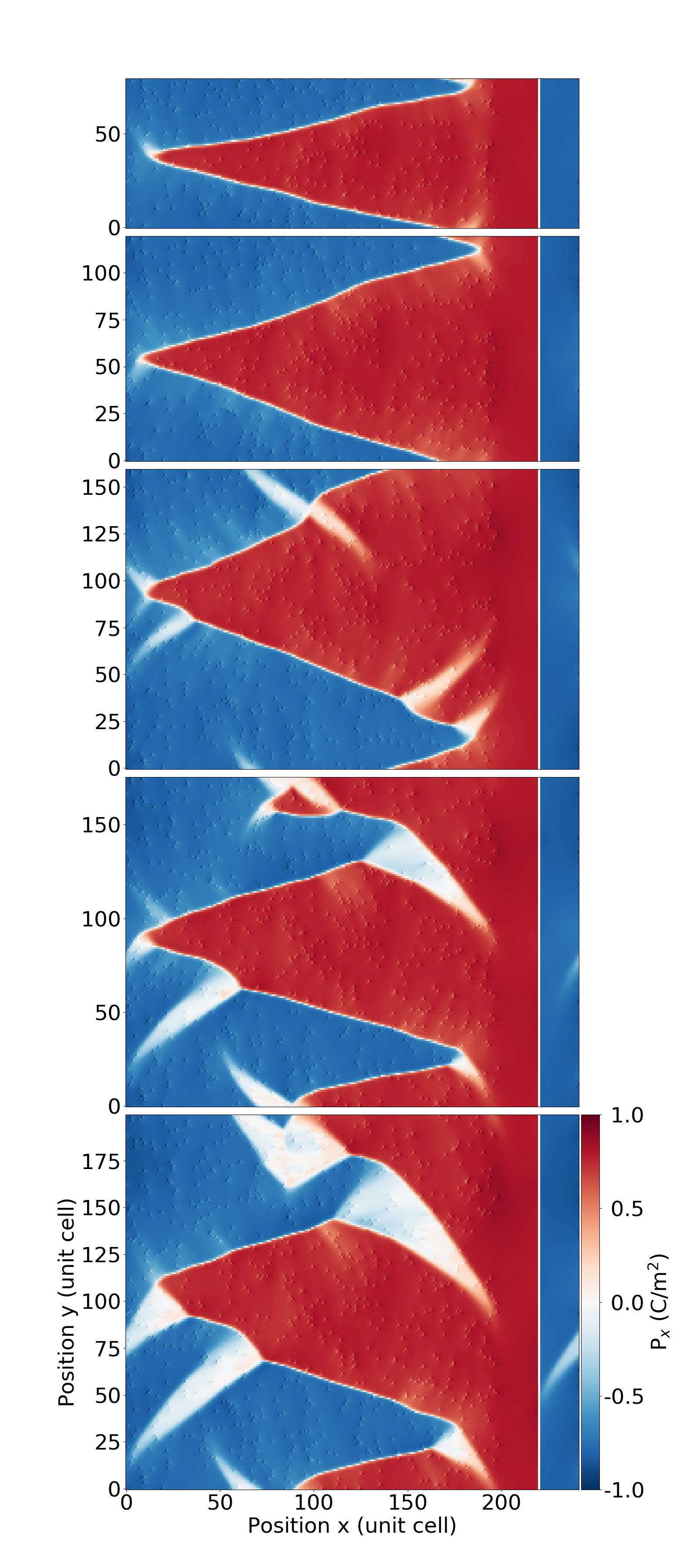}
\caption{Zig-zag domain wall as obtained from mutually independent phase-field simulations for $L=200$\,$\Delta$ and  $W\in\{80,120,160,176,200\}\Delta$. The color represents $P_\mathrm{x}$.  White regions are areas where the polarization is oriented along the $y$-axis.}
\label{fig_large_L_changing_W}
\end{figure}

As it was already pointed out, there appear a smoke-like features at the tips of the triangles (see Fig.\,\ref{fig_shell_model_pyramid_mauro1}(b) and Fig.\,\ref{fig_phase-field_diffent_charges}(b)). These features are related to the somewhat smaller height of the wedge compared to the width of the charged region $L$, and they also develop as a response to the discrepancy between $W_\mathrm{natural}(L)$ and the actual $W$ (enforced by the $y$-dimension of the simulation box), in particular when $W$ is larger. To demonstrate this, we show an even more pronounced departure from the ideal zig-zag pattern in Fig.\,\ref{fig_large_L_changing_W}, observed for triangles with increasing base $W$.
The white regions in the figure represent small 90$^\circ$ domains with the ferroelectric polarization pointing along the $\pm y$-direction. Notice that the resulting 90$^\circ$ domain walls tend to form an angle of 45$^\circ$ with respect to the pseudocubic axes, i.e. they are close to mechanical compatibility.\cite{fousek:1969:orientation}
 
For values of $W$ smaller than $W_\mathrm{natural}$ the shape remains triangular, even though the overall wall length and hence its surface energy increase. This is because failure of the wall to percolate the complete thickness of the charged slab would be energetically forbidden, as explained above. 

For values of $W$ larger than $W_\mathrm{natural}$ the shape departs from triangular only very slowly, indicating a relatively large tolerance of the system to the change of the triangle width. At some point, the smoke-like feature above the triangle grows larger, and is accompanied by smaller and larger cuts to the side of the triangle along its whole length. The actual positions of the cuts likely depend on the location of defect charges. For large $W$, the cut progressively separates the bottom part of the triangle, which becomes a seed for a second triangle, which will fully develop into another triangular domain for large enough $W$.

Let us point out that in a real situation the width of the triangles will likely depend on the history of the sample, and might be far from optimal. It will depend on, e.g., the number of seeds during the growth of the domain structure, on any pinning of the wall (influencing the ability of the triangles to move, merge or split), and on other aspects of the charged-domain wall formation. Thus, an irregular pattern and varying heights of the triangles can be expected.


The other situation for which the wall morphology strongly departs from the regular zig-zag pattern appears for very small $L\lesssim 10\Delta$. Here the domain wall forms an irregular landscape within the limits of the charged region, utilizing the energetical advantage of crossing the oppositely charged defect in the wall center, if possible. We observed that in this case the wall is pinned to the actual defect positions.
%
The absence of the regular zig-zag pattern is the reason why it was not possible to plot W$_\mathrm{natural}$(L) for small $L$ in Fig.\,\ref{fig_W_natural}.

\subsection{Experimental context} 

An experimental observation of a similar configuration of charged domain wall was recently reported in Ref.\,\onlinecite{art_denneulin_2022} for a 250\,nm thick BaTiO$_3$ film. Therein, a charged wall separates two regions with oppositely oriented polarization (perpendicular to the substrate). Triangular domains, which penetrate almost the entire film thickness, are observed along with smaller triangles. A similar pattern was also observed in rhombohedral BiFeO$_3$ in Ref.\,\onlinecite{art_jia_2015}. The structures of these walls bear a strong resemblance to the zig-zag pattern studied here (we are aware that formation of a zig-zag pattern was addressed theoretically in Ref.\,\onlinecite{art_zhang_2020}, without the necessity to consider broad distribution of compensation charges).
A triangular charged domain interface (on much larger scale) was also observed 
for LiNbO$_3$.\cite{art_schroder_2012}

Zig-zag charged
domain walls similar to those considered here are known to appear in magnetic thin films on scales of micrometers and larger.\cite{art_hamzaoui_1984,art_engel_2006,art_favieres_2020,art_knupfer_2021} Some of the triangles show features similar smoke-like features as those described here. The compensating electric charge used in the electrostatic considerations performed here is analogous to the notion of the magnetic charge due to orientation of the magnetization out of the plane of the film. In this way, the magnetic charged wall can exist in thin films, while the out of plane component of the magnetization escapes the film and the magnetic flux closes outside the film.

\section{Conclusions} 

We used atomistic shell-model and continuous phase-field simulations of doped ferroelectric PbTiO$_3$ to study the properties of 180$^\circ$ tail-to-tail domain walls that develop in spatially extended charge-compensation layers.

We observe that the charged domain walls systematically adopt a zig-zag profile. We show that this pattern is stable against variations of compensating charge distribution and forms equally for point-charges of different magnitude and homogeneous charge distributions.
We argue that the zig-zag shape and triangular domains form as a consequence of the energetic demand to compensate the charged layer via polarization gradients while avoiding the paraelectric state, and keeping the surface area of the wall as small as possible.
The former is achieved by polarization rotation. Additionally, we provide a simplified expression for determining the natural width of the triangles $W_\mathrm{natural}$, or equivalently the angle of the ziz-zag domain wall. For large enough thickness $L$ of the charged slabs we found $W_\mathrm{natural}\approx L^{2/3}$.


Although the charged region considered here was a slab with the normal aligned with the direction of spontaneous polarization in the surrounding domains, the physics presented here can be straightforwardly extended to more general orientations of the slab with respect to spontaneous polarization, and to non-180$^\circ$ charged domain walls. 
%
This will be relevant, for example, for the slab normal along the pseudocubic axis in a rhombohedral ferroelectrics, such as in BiFeO$_3$.
%

The results presented here will be equally valid for the positively charged head-to-head wall, provided the compensation-defect charges remain distributed in broad region and are immobile.


\section{Acknowledgement} 

This work was supported by the Czech Science Foundation (Project No. 20-05167Y).
%
Computational resources were supplied by the project "e-Infrastruktura CZ" (e-INFRA CZ LM2018140 ) supported by the Ministry of Education, Youth and Sports of the Czech Republic.

\bibliographystyle{apsrev}
\bibliography{article_all}

\end{document}